\documentclass[11pt,twoside]{article}

\usepackage{asp2006}
\usepackage{epsf}
\usepackage{psfig}
\usepackage{lscape}
\usepackage{epsfig}

\begin{document}

\title{The WiggleZ project: AAOmega and Dark Energy}

\author{Karl Glazebrook, Chris Blake, Warrick Couch, Duncan Forbes\altaffilmark{1}, Michael Drinkwater, Russell Jurek, Kevin Pimbblet\altaffilmark{2}, Barry Madore\altaffilmark{3}, Chris Martin, Todd Small, Karl Forster\altaffilmark{4}, Matthew Colless, Rob Sharp\altaffilmark{5}, Scott Croom\altaffilmark{6}, David Woods\altaffilmark{7}, Michael Pracy\altaffilmark{8}, David Gilbank, Howard Yee\altaffilmark{9}, and Mike Gladders\altaffilmark{10}}

\affil{}

\altaffiltext{1}{Centre for Astrophysics \& Supercomputing, Swinburne
  University of Technology, P.O. Box 218, Hawthorn, VIC 3122,
  Australia}

\altaffiltext{2}{Department of Physics, University of Queensland,
  Brisbane, QLD 4072, Australia}

\altaffiltext{3}{Observatories of the Carnegie Institution of
  Washington, 813 Santa Barbara Street, Pasadena, CA 91101, USA}

\altaffiltext{4}{California Institute of Technology, MC 405-47, 1200
  East California Blvd., Pasadena, CA 91125, USA}

\altaffiltext{5}{Anglo-Australian Observatory, PO Box 296, Epping, NSW
  1710, Australia}

\altaffiltext{6}{School of Physics, University of Sydney, Sydney, NSW
  2006, Australia}

\altaffiltext{7}{Department of Astrophysics, University of New South
  Wales, Sydney, NSW 2052, Australia}

\altaffiltext{8}{Research School of Astronomy and Astrophysics,
  Australian National University, Canberra, ACT 2600, Australia}

\altaffiltext{9}{Department of Astronomy \& Astrophysics, University
  of Toronto, 50 St. George Street, Toronto, ON, M5S 3H4, Canada}

\altaffiltext{10}{Department of Astronomy and Astrophysics, University
  of Chicago, 5640 South Ellis Avenue, Chicago, IL 60637, USA}

\begin{abstract}

We describe the `WiggleZ' spectroscopic survey of 280,000 star-forming
galaxies selected from a combination of GALEX ultra-violet and SDSS $+$ RCS2
optical imaging. The fundamental goal is a detection of the baryonic
acoustic oscillations in galaxy clustering at high-redshift ($0.5 < z
< 1$) and a precise measurement of the equation of state of dark
energy from this purely geometric and robust method.  The survey has
already started on the 3.9m Anglo-Australian Telescope using the AAOmega
spectrograph, and planned to complete during 2009. The WWW page for
the survey can be found at {\tt astronomy.swin.edu.au/wigglez}. \\ \\
{\bf Version 2: updated May 2007 with final analysis of data through to Nov 2006.}

\end{abstract}

\section{Introduction}

One of the major triumphs of modern astrophysics over the last decade
has been the extraordinary precision with which the cosmological parameters
can now be measured from new experiments.
The prime examples are measurements
of temperature anisotropies in the Cosmic Microwave Background (CMB)
radiation, the clustering of galaxies on large scales, the maximum
brightness of distant Type Ia supernovae (SN), and the primordial
abundances of the light elements. The age, expansion rate, geometry,
matter and energy content of the Universe can now be determined to a
precision of better than 10\%.  The remaining
challenge to understanding the underlying physics is the profound
discovery from distant SN studies that the expansion rate of the
Universe is accelerating (Riess et al.\ 1998, Perlmutter et
al.\ 1999). This result  implies
that the cosmic energy budget must be dominated by a new form of
matter which has a negative pressure -- `dark energy' (Deffayet et
al.\ 2002). The simplest explanation is an inherent energy density of
the quantum vacuum (a `cosmological constant' term) but is 
far short (by a factor of $c^5\, G^{-1} \hbar^{-1} H_0^{-2}$ $\sim 10^{122}$)
of the natural  Planck energy density, thus motivating
alternative models for the
dark energy.
Understanding the nature of this dark energy is one of the key
`Science Questions for the New Century' (Turner et al.\ 2003).

The most direct existing measurement of the cosmic acceleration comes
from the observation of distant SN as `standard candles'.
After correction for the luminosity-light curve width correlation,
these SN allow the measurement of distances to redshifts as
high as $z = 1.7$ (Riess et al.\ 2004).  However, there are concerns
over potential systematic evolutionary effects masquerading as
cosmological effects, suggesting that an independent cross-check of
this result is crucial.  A conceptually similar approach would be to
use a `standard ruler', or cosmic feature with a known absolute
length-scale.  The apparent size of this feature at a given redshift
would yield the cosmic distance to that redshift.  Since individual
objects evolve, accurate cosmic measuring rods are rare.  However, a
newly-developed method is to use `Baryonic Acoustic Oscillations'
(BAO), which are features imprinted in the power spectrum of galaxy
clustering. These features arise from acoustic oscillations in the
dense early Universe that have a preferred scale, the `sound
horizon', and get frozen into the distribution of matter after
recombination.  After galaxy formation they are well-preserved on
large scales despite the growth of non-linearities on smaller scales
(Eisenstein, Seo \& White 2007). Since the size of the sound horizon
at recombination is accurately calibrated by CMB measurements, these
`baryon wiggles' act as a cosmic standard ruler and can in principle
be used as a probe of dark energy (Blake \& Glazebrook 2003, Seo \&
Eisenstein 2003).  Currently the BAOs are only well-detected in the
local Universe (Eisenstein et al.\ 2005 (E05), Cole et al.\ 2005), but in
principle BAO galaxy redshift surveys can be performed at high
redshift to measure precise distance scales.  Spectroscopic surveys of
$\ga 10^3$ deg$^2$ and $\ga 10^5$ galaxies are fundamentally required
(Glazebrook \& Blake 2005).

\vfill

\section{The WiggleZ survey}

The goal of the `WiggleZ' survey is to achieve the first detection of
BAO features at {\it high-redshift\/} in a spectroscopic survey.  A
spectroscopic BAO survey has several requirements.  A wide instrument
field-of-view is desirable to cover large areas efficiently.  A
large-aperture telescope is required to reach high redshifts. A ready
supply of deep well-calibrated imaging is necessary -- which is one of
the more difficult requirements to meet.  One flexibility we can utilize 
is to sparsely-sample the large-scale structure with whichever 
class of galaxy allows the most rapid measurement of
redshifts, given
that all types of galaxies possess an approximately linear clustering
bias on large scales.  Also, the required redshift resolution is modest
($\Delta z \la 0.001$), demanding only low-resolution spectroscopy.
For the WiggleZ survey we found a good match to all these requirements
from the combination of the wide-field ultra-violet GALEX satellite,
the Sloan Digital Sky Survey (SDSS) and the AAOmega spectrograph on
the 3.9m Anglo-Australian Telescope (AAT).  This has allowed us to recently
start a $0.5<z<1.0$ BAO survey, targetting 400,000 objects (yielding 280,000 
redshifts and 210,000 with $z>0.5$), over
1000 deg$^2$, which is planned to be completed during 2009.

We note that photometric-redshift surveys have already produced a tentative
detection of BAOs at $z \approx 0.5$ (Padmanabhan et al.\ 2007, Blake
et al.\ 2007).  There are two main disadvantages of photo-$z$ BAO
surveys.  Firstly, the radial smearing due to the photo-$z$ error
reduces the signal-to-noise of the power spectrum measurement.
Secondly, spectroscopic surveys can resolve the oscillations in the
{\it radial} and tangential directions, allowing
direct determinations of the Hubble expansion rate $H(z)$ as well as
the more conventional metric distance $D_A(z)$.  Measurement of $H(z)$
at high-$z$ is a unique feature of the BAO method.

\subsection{Survey design: Imaging}

When designing a BAO survey at high redshift we faced a key initial
choice: {\it red galaxies or blue galaxies?}  Red galaxies possess a
higher clustering strength (galaxy bias), implying that a lower number
density is required to minimize the shot noise of the large-scale
structure measurement.  Indeed, Luminous Red Galaxies in the SDSS
produced the existing low-$z$ BAO detection.  On the other hand, it
requires significantly more exposure time to obtain redshifts for red
galaxies owing to their lack of emission lines and the consequent need
to detect the galaxy continuum with a reasonable signal-to-noise
ratio.  Moreover, the higher clustering amplitude of red galaxies
implies that the non-linear growth of structure (which erases the
BAOs) influences the clustering pattern on larger scales.  Our
calculations of these factors showed that the blue galaxies were
slightly superior, with the emission-line redshifts promising greater
robustness in poor observing conditions.  Furthermore, we identified a
clean way to select these blue galaxies as discussed below.  We
suggest that it will prove extremely valuable to pursue BAO detections
using both blue and red galaxy populations, in order to understand the
subtle systematic effects due to galaxy formation processes, and
discover if these effects produce any significant distortion of the
acoustic oscillation signature.  We note that there are several
proposed new surveys of red galaxies for BAO measurements in the $z<1$
regime (Eisenstein et al., these proceedings), which the WiggleZ
survey will complement.

A very elegant method of selecting emission-line galaxies with
redshifts $z>0.5$ is to use ultra-violet data from the GALEX satellite
(Martin et al.\ 2005). Star-forming galaxies with $z > 0.5$ drop out
of the far-UV (FUV) filter, due to the Lyman break, but have blue
near-UV (NUV) to optical colours.  We select galaxies with primary criteria of
$20 < r <22.5$, $NUV < 22.8$, $FUV-NUV > 1$ and $NUV - r < 2$. Analysis of
the early data taken (through Nov 2006) shows that a secondary
$(g-r, r-i)$ optical colour cut applied at bright magnitudes significantly improves the removal of $z<0.5$ foreground
objects and this will be used in future telescope runs. The final result is a target catalog of 400
spectroscopic targets deg$^{-2}$. These are luminous star-forming galaxies from the tip of the galaxy luminosity
function at $z \sim 0.7$ analogous to (though somewhat redder than) the Lyman Break Galaxies at
$z \sim 3$ (Steidel et al.\ 1996).  The optical data is also required
to produce an accurate fibre position for spectroscopy.  The UV limit
 is the depth of the GALEX Medium Imaging Survey (MIS).
The final galaxy density of 210 deg$^{-2}$ is calculated as sufficient for
suppressing galaxy shot noise assuming a reasonable clustering bias
for high-redshift star-forming galaxies and the observed completeness.

\begin{figure}[t]
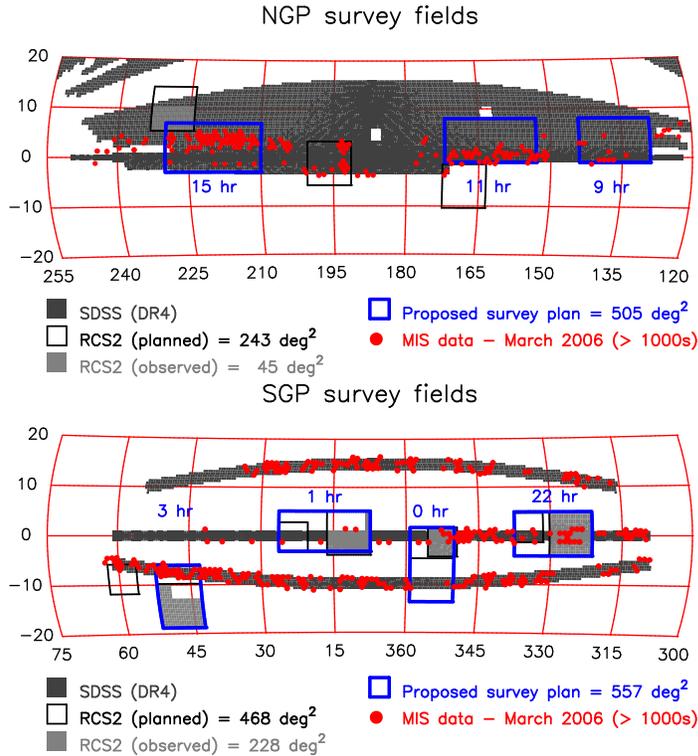

\begin{center}
\epsfig{file=ngp_plan_all.ps,width=5cm,angle=-90}
\epsfig{file=sgp_plan_all.ps,width=5cm,angle=-90}
\end{center}
 \caption{Layout of the WiggleZ fields in the NGP and SGP regions,
   illustrating the available SDSS, RCS2 and GALEX data.}
\end{figure}

The choice of fields (Figure~1) is driven by the availability of
existing optical data and the requirement that the sky patches be
large and square enough (at least $\sim 9\deg$ on each side) to probe
structure on the scales relevant for BAOs (i.e., at least 2-3 times
larger than the standard ruler size of $105 \, h^{-1}$ Mpc).  In the
North Galactic Cap we are able to utilize the existing Sloan Digital
Sky Survey (York et al.\ 2000). In the south, the SDSS stripes are too
narrow (Figure~1).  Accordingly, we supplement SDSS with data from the
RCS2 survey (Yee et al., these proceedings) supplied to us via a
collaborative arrangement.  Additionally, we have chosen fields where
large amounts of GALEX MIS data already exist.  However, these data are
patchy owing to GALEX bright-star avoidance constraints.  This
patchiness is on the scale of the GALEX field-of-view ($\approx
1\deg$) and would produce unacceptable convolution of our measured
power spectrum and damping of the measured BAOs.  We are therefore
using GALEX to tile these fields with a higher filling factor ($\ga
70\%$, determined from simulations). This requires a modest additional
amount of GALEX time at the rate of 400 orbits per year to extend MIS in order to
feed the spectroscopic pipeline.  Some care is needed with the GALEX
observing plan: `petal pattern' pointings must be used in some areas
containing moderately bright stars.

\subsection{Survey design: Spectroscopy}

Sources are selected from matched UV and optical catalogues, and
allocated to AAOmega pointing centres using a simulated annealing
algorithm (Campbell, Saunders \& Colless 2004).  Fainter optical
sources at preferentially higher redshift are assigned higher priority
in the annealing scheme.  On average, each part of the sky in our
survey must be visited 4 times.

\begin{figure}[t]
\begin{minipage}{3.5cm}Figure~2.  The galaxy redshift distribution resulting from the current
  WiggleZ survey target selection strategy (data through Nov 2006) compared to the 2dF Galaxy Redshift Survey ($z\sim 0.1$).
\end{minipage}
\hspace{5mm}
\begin{minipage}{3.5in}
\psfig{file=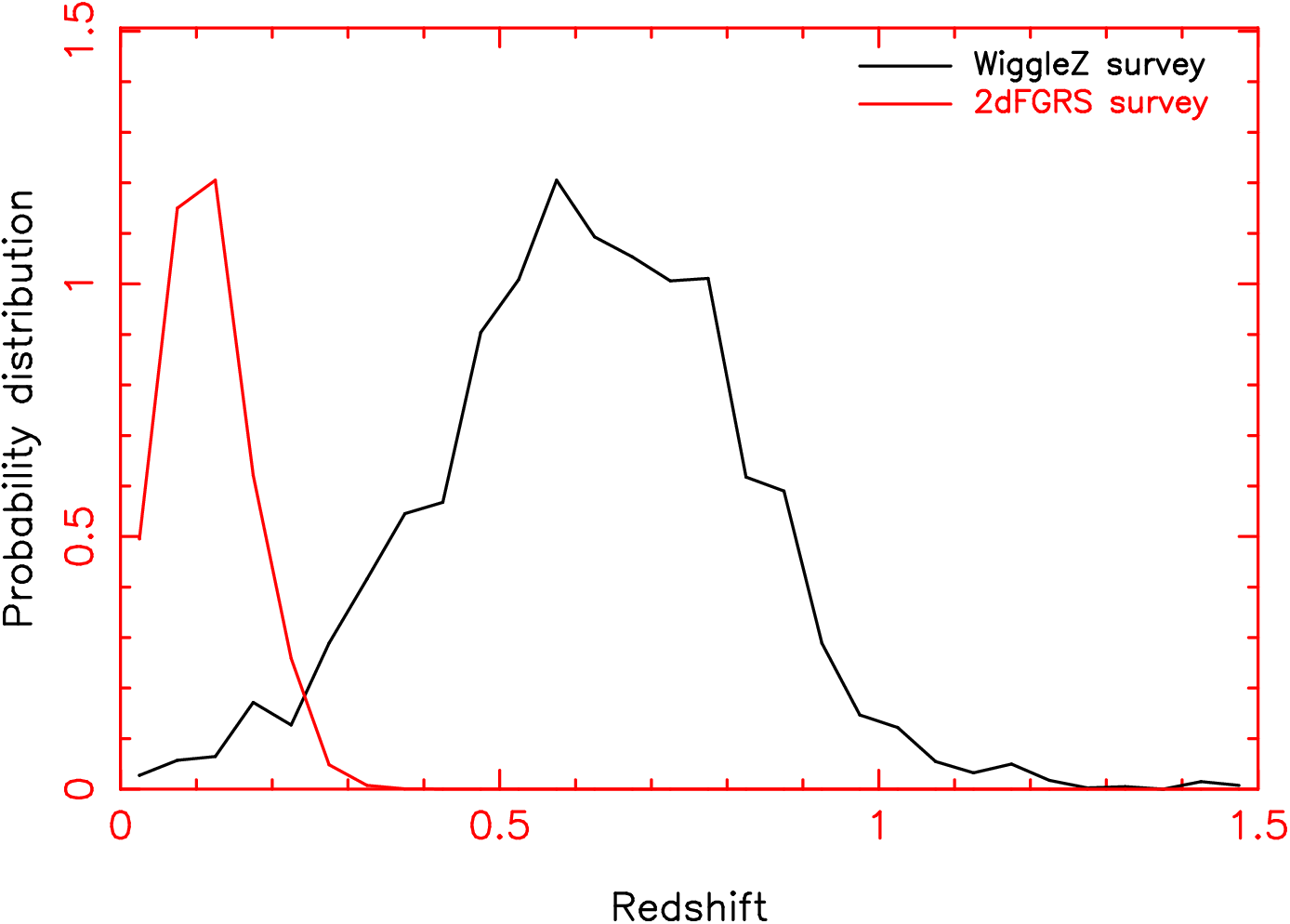,width=6.0cm,angle=270}
\end{minipage}
\end{figure}

The WiggleZ survey is performed using the AAOmega spectrograph at the
AAT.  AAOmega is the upgraded Two Degree Field
(2dF) system and is capable of taking spectra of 400 objects
simultaneously across a 2$\deg$ diameter field-of-view using optical
fibres positioned by an $XYZ$ robot. The fibres run down to a
dual-beam spectrograph in the Coud\'e room which (for our setup)
covers 3700--8750\AA\ at a spectroscopic resolution of 5\AA\ (FWHM).
A full description of AAOmega and its various modes is given by Sharp
et al.\ (2006).

One hour exposures with AAOmega yield redshifts for 70\% of our
galaxies in average usable conditions, for our optimized target
selection.  1 hr is well-matched to the fibre reconfiguration time of
the positioner system. Longer exposures would result in higher completeness
but would also result in less area covered per unit telescope time and this trade is
not beneficial.
Redshifts are usually identified from the
[OII] and H$\beta$/[OIII] emission lines. Continuum signal:noise is low.
For $z>0.8$ only [OII] is
visible, but at our spectroscopic resolution it is usually marginally
resolved: we either see two peaks from the doublet or a `fat'
line. The only possible sources of contamination of emission lines
with similar equivalent widths are H$\alpha$ at $z\sim 0.1$ (removed
by our FUV$-$NUV selection), or Ly$\alpha$ at $z\sim 5$ (which would
have no UV flux).  Additionally H$\alpha$ would have to be very low
metallicity to avoid [NII] being picked up.  Thus we believe that our
single-line redshift identifications are robust.

The resulting galaxy redshift distribution is shown in Figure~2 and
covers the range $z < 1.3$.  For the baryon oscillations analysis we
plan to focus on redshift bins in the range $z > 0.5$.  The entire
redshift range will be used for galaxy evolution studies.  Currently,
about $75\%$ of redshifts satisfy $z > 0.5$ with our most optimum colour
cuts.  The
ability to reach up to $z \sim 1$ in the same exposure time as the
previous 2dF Galaxy Redshift Survey at $z \sim 0.1$ (Colless et
al.\ 2001) is due to a combination of increased spectrograph
sensitivity and the judicious target selection.

\section{Current Status and Predicted Survey Performance}

\begin{figure}[t]
\begin{minipage}{3.5cm}Figure~3.  The projected real-space correlation function of 5849
 WiggleZ survey galaxies (Nov 2006 data) in three fields.  A spatial correlation function
  $\xi(r) = (r/r_0)^{-\gamma}$ was assumed in the fit.
\end{minipage}
\hspace{5mm}
\begin{minipage}{3.5in}
\psfig{file=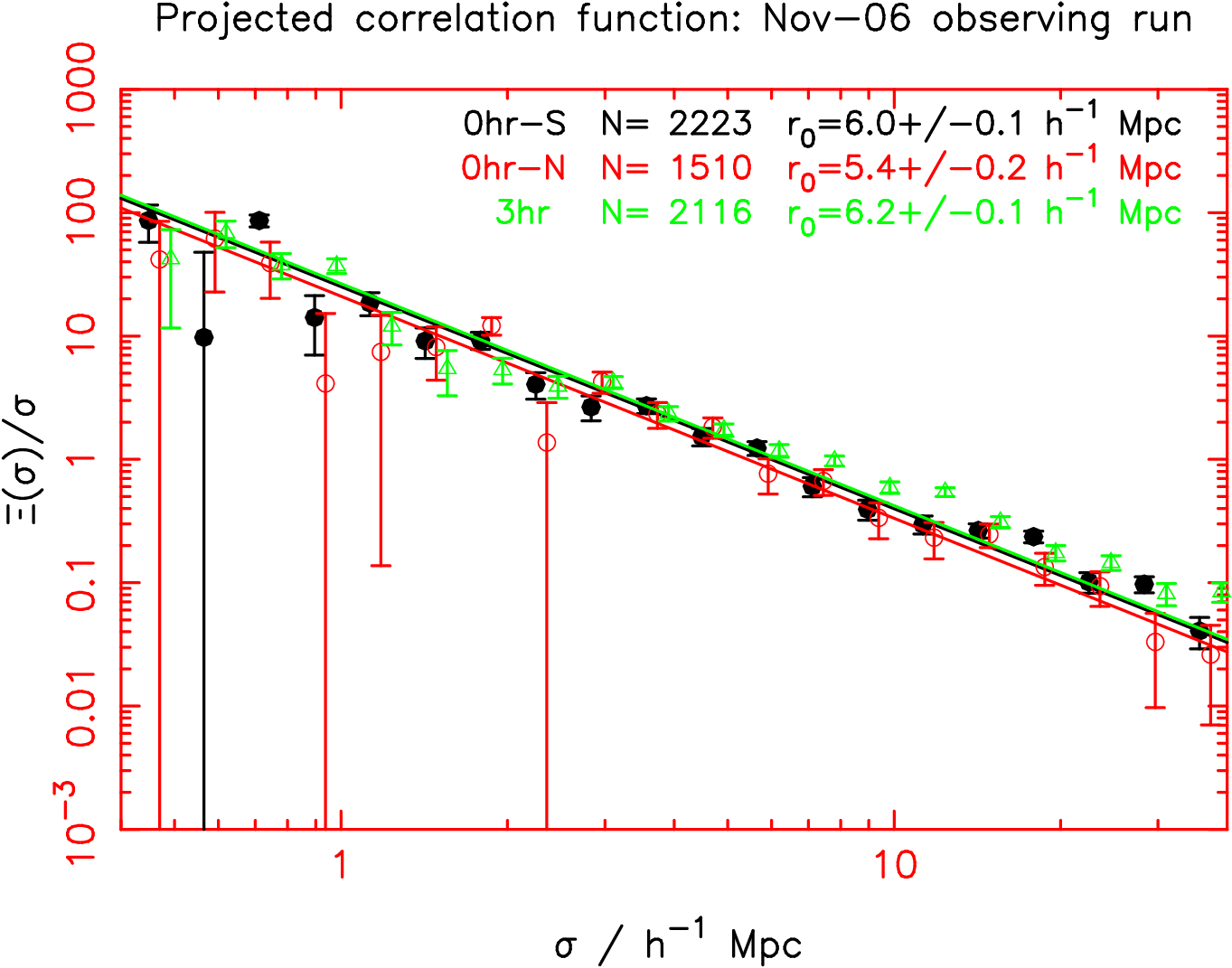,width=6.0cm,angle=270}
\end{minipage}
\end{figure}

There have been two WiggleZ survey observing runs in August and
November 2006, during which we accumulated a combined total of
$15{,}000$ redshifts.  GALEX data acquisition is on-going. The full
220 nights of AAT time has been awarded, subject to a mid-term review
in 2008.  The WiggleZ team is planning mid-term and final public data
releases in 2008 and 2009 respectively, in a VO-compliant format
together with an SQL database.

We have measured the clustering properties of  our sample using 
5849 redshifts obtained to date. The preliminary measurement of the
two-point correlation function (Figure~3) is consistent with other measurements of 
the most luminous
blue galaxies at these redshifts (Coil et al.\ 2006, Meneux et
al.\ 2006).  They are more clustered than typical blue galaxies because
of the low space density and higher luminosity ($L>L^*_{UV}$). 
 
The clustering length is about $r_0\sim 6 h^{-1}$Mpc in real space 
(7 $h^{-1}$Mpc in redshift space). From this 
 we can predict the
signal-to-noise ratio of the final power spectrum measurement for a
1000 deg$^2$ survey.  A simulation of this is shown in Figure~4.  It
can be seen that the sinusoidal BAO features are well detected; a
Monte-Carlo analysis (Blake \& Glazebrook 2003) finds that the BAO
scale (and consequently the standard ruler at $\overline z \sim 0.7$)
can be measured with an accuracy of about $2\%$.  A full analysis of
radial and tangential Fourier modes predicts that we can measure the
expansion rate of the Universe at high redshift, $H(z=0.7)$, with 4\%
accuracy.  Implementing the improved method of `reconstructing' the
density field (Eisenstein et al.\ 2007), the predicted measurements of
$D_A$ and $H$ are $1.8\%$ and $2.7\%$, respectively.

\begin{figure}[t]
\begin{minipage}{3.5cm}Figure~4.  Simulated final survey power spectrum $P(k)$ of the WiggleZ
  survey.  The overall shape of the power spectrum has been divided
  out to highlight the BAO features.
\end{minipage}
\hspace{5mm}
\begin{minipage}{3.5in}
\psfig{file=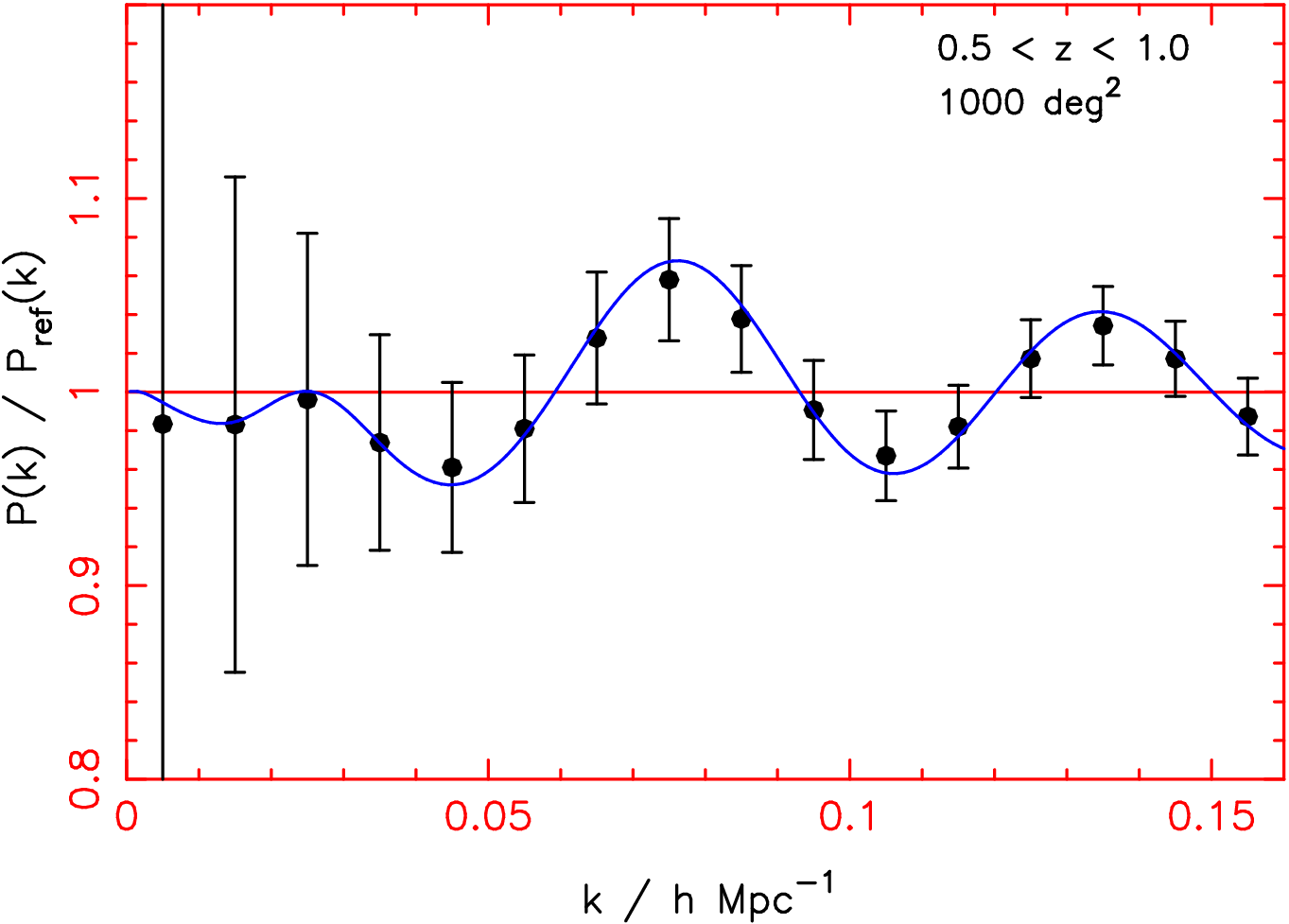,width=5.0cm,angle=270}
\end{minipage}
\end{figure}

These distance measurements can be propagated to constraints on the
cosmological parameters, including the properties of dark energy,
characterizing it by an equation of state $w = P/\rho$ (where $w
= -1$ for a pure cosmological constant).  Essentially we measure the
cosmic distance and expansion rate at $z \approx 0.7$ in units of the
sound horizon at recombination.  In addition to $w$, these quantities
depend on the matter density $\Omega_{\rm m}$, the zero redshift
Hubble parameter $h$, and potentially a non-zero curvature
$\Omega_{\rm k}$.  Independent measures of these additional parameters
are possible using the CMB or overall shape of the galaxy power
spectrum.  As an example, Figure~5 shows simulated parameter
measurements in the $(\Omega_{\rm m},w)$ space, assuming
$\Omega_{\rm k} = 0$ and an external prior $\sigma(\Omega_{\rm m}) =
0.03$.  Confidence contours for the WiggleZ survey (assuming the
conservative case of no `density reconstruction') are compared to
those from the Supernova Legacy Survey current data (Astier et al.\ 2006)
in order to illustrate the superb complementarity and systematic
cross-checking possible with these two probes of dark energy.  The
equation of state can be measured with accuracy $\Delta w \approx 0.1$
in both cases.  Introducing a curvature degree of freedom does not
degrade these contours significantly, because curvature is tightly
constrained by these distance measurements in conjunction with the
CMB.  In this context the BAO and SN methods are also
complementary, because the BAOs measure distances relative to the
last-scattering surface at $z \approx 1100$, whilst SN measure
distances relative to a local calibration at $z \approx 0$.

In addition to the dark energy measurements, such a large galaxy
spectroscopic survey will deliver much more science.  A partial list
includes: tests of neutrino masses and the physics of inflation using
the shape of the large-scale power spectrum; testing galaxy formation
models via small-scale clustering measurements and the dependence of
luminosity functions, star-formation rates, metallicity and colour on
environment.
 
\section{Summary}
 
The WiggleZ survey is on schedule to deliver the first precision
high-redshift BAO dark energy constraints. It will be an independent,
complementary and constraining test of the accelerating Universe
paradigm: measuring cosmological acceleration completely independently of any
SN systematic effects; complementary to SN
observations via orthogonal confidence regions; and highly
constraining by virtue of per-cent level distance and expansion rate
measurements.  Given the current progress of the WiggleZ survey we are
expecting our first BAO detection and dark energy results to appear in
2008. The results from WiggleZ will have a major impact on the design
of future BAO experiments using new instruments and larger 8m-class
telescopes, such as WFMOS (Glazebrook et al.\ 2005) and HETDEX (Hill et
al.\ 2004).

 \begin{figure}[t]
\begin{minipage}{4.5cm}Figure~5.  Simulated cosmological parameter measurements in the space of
  $(\Omega_{\rm m},w)$ comparing those from 
   the future WiggleZ survey ($+$ the E05  $z\simeq 0.35$ BAO constraint from SDSS) 
   with  the current Supernova
  Legacy Survey data. 
\end{minipage}
\hspace{5mm}
\begin{minipage}{3.5in}
\psfig{file=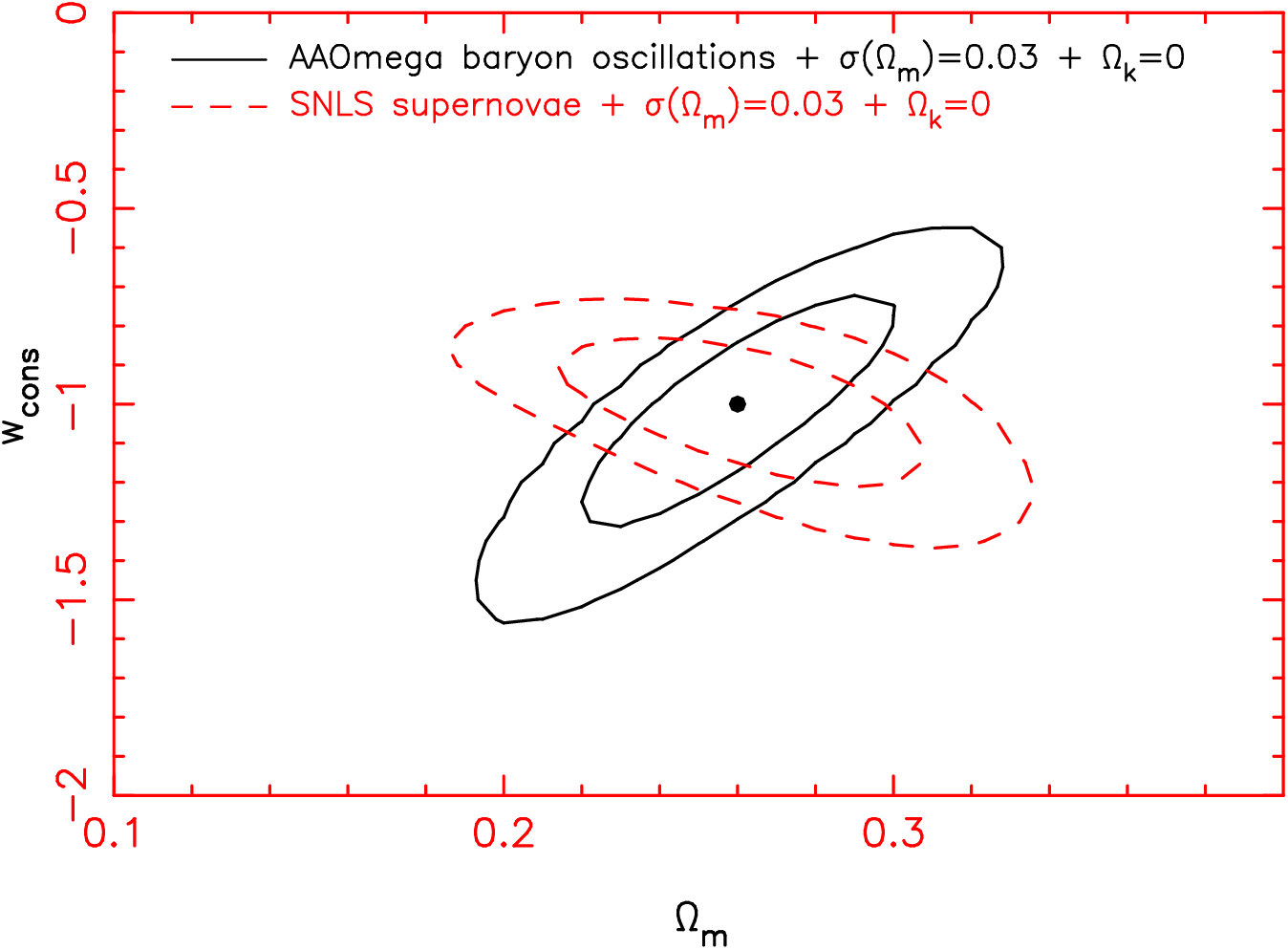,width=5.0cm,angle=270}
\end{minipage}
\end{figure}

\acknowledgements WiggleZ is supported by a generous allocation of
AAT time and an Australian Research Council
Discovery Project Grant \#DP0772084.  GALEX is a NASA small explorer
launched in April 2003. We gratefully acknowledge NASA's support for
construction, operation, and science analysis for the GALEX mission,
developed in cooperation with the Centre National d'\'Etudes
Spatiales of France and the Korean Ministry of Science and Technology.


\begin{thebibliography}{}

\bibitem{1} Astier P., et al., 2006, \aap, 447, 31

\bibitem{2} Blake C., Glazebrook K., 2003, ApJ, 594, 665

\bibitem{3} Blake C., Collister A., Bridle S., Lahav O., 2007, MNRAS,
  in press (astro-ph/0605303)

\bibitem{4} Campbell L., Saunders W., Colless M., 2004, MNRAS, 350,
  1467

\bibitem{5} Coil A., Newman J., Cooper M., Davis M., Faber S., Koo
  D., Willmer C., 2006, \apj, 644, 671

\bibitem{6} Cole S., et al., 2005, MNRAS, 362, 505

\bibitem{7} Colless M., et al., 2001, \mnras, 328, 1039

\bibitem{8} Deffayet C., et al., 2002, PRD, 65, 044023

\bibitem{9} Eisenstein D., et al., 2005, ApJ, 633, 560 (E05)

\bibitem{10} Eisenstein D., Seo H.-J., White M., 2007, ApJ, in press
  (astro-ph/0604361)

\bibitem{11} Eisenstein D., Seo H.-J., Sirko E., Spergel D., 2007,
  ApJ, in press (astro-ph/0604362)

\bibitem{13} Glazebrook K., Blake C., 2005, ApJ, 631, 1

\bibitem{14} Glazebrook K., Eisenstein D., Dey A., Nichol B., \& The
  WFMOS Feasibility Study Dark Energy Team, 2005, Dark Energy Task
  Force White Paper (astro-ph/0507457)

\bibitem{12} Hill G., Gebhardt K., Komatsu E., MacQueen P., 2004, AIP
  Conf.~Proc.~743: The New Cosmology: Conference on Strings and
  Cosmology, 743, 224

\bibitem{15} Martin D., et al., 2005, ApJ, 619, L1

\bibitem{16} Meneux B., et al., 2006, \aap, 452, 387

\bibitem{17} Padmanabhan N., et al., 2007, MNRAS, in press
  (astro-ph/0605302)

\bibitem{20} Perlmutter S., et al., 1999, ApJ, 517, 565

\bibitem{18} Riess A., et al., 1998, AJ, 116, 1009

\bibitem{19} Riess A., et al., 2004, ApJ, 607, 665

\bibitem{21} Seo H.-J., Eisenstein D., 2003, ApJ, 598, 720

\bibitem{22} Sharp et al., 2006, SPIE, 6269, 14

\bibitem{23} Steidel C., Giavalisco M., Pettini M., Dickinson M.,
  Adelberger K., 1996, \apjl, 462, L17

\bibitem{24} Turner M., et al., 2003, {\it Connecting Quarks with
  the Cosmos: Eleven Science Questions for the New Century}, National
  Academies Press: Washington

\bibitem{25} York D., et al., 2000, \aj, 120, 1579

\end{thebibliography}
\end{document}